\documentstyle[aps,preprint,floats,epsf,epsfig]{revtex}
\begin{document}

\tighten

\title{A Lattice Study Of The Magnetic Moment \\
And The Spin Structure Of The Nucleon
\footnote{This work is based on V.G.'s research for a dissertation 
to be submitted to the Graduate School, University of Maryland, by V.G. in partial fulfillment
of the requirement for the PhD degree in physics.}  }

\bigskip

\author{Valeriya Gadiyak, Xiangdong Ji, Chulwoo Jung}

\address{ Department of Physics \\
University of Maryland \\
College Park, Maryland 20742 } 

\maketitle

\begin{abstract}

Using an approach free from momentum 
extrapolation, we calculate the nucleon magnetic moment and the 
fraction of the nucleon spin carried by the quark angular momentum
in the quenched lattice QCD approximation. 
Quarks with three values of lattice masses,
$210,~ 124$ and $80~ {\rm MeV}$, are formulated on 
the lattice using the 
standard Wilson approach. At every mass, 100 gluon 
configurations on $16^3\times 32$ lattice 
with $\beta=6.0$ are used for statistical averaging. 
The results are compared 
with the previous calculations with momentum extrapolation. 
The contribution of the disconnected diagrams 
is studied at the largest quark mass using noise 
theory technique. 

\end{abstract}

\narrowtext

\newpage

\section{Introduction} 

The magnetic moment is one of the fundamental properties of the
nucleon. Experimentally this quantity has been measured to 
a very high precision \cite{EPJ}. In the nucleon model-building, 
the magnetic moment is usually the first to check against the 
experimental data. To understand how well a particular version
of lattice QCD can simulate the internal structure of the 
nucleon, the magnetic moment is an observable that one 
must study after the nucleon mass.

Theoretical computations of the nucleon magnetic moment 
using lattice QCD techniques have been performed by 
a number of groups on lattices of different physical 
sizes and lattice spacings with different nucleon 
sources (interpolating fields) \cite{draper}-\cite{williams}. 
Most calculations done so far have used the quenched 
approximation. And almost all calculations begin 
with the lattice calculations of Sach's magnetic 
form-factor $G_M(Q^2)$, where $Q^2 = -q_0^2 + {\vec q}^2$, 
at discrete values of the momentum transfer $\vec q$. 
The magnetic moment is just $G_M(Q^2)$ at $Q^2=0$
which can not be obtained more directly in this approach  
because the relevant off-forward matrix element 
is linearly proportional to ${\vec q}$.
The lattice magnetic momentum is obtained usually by 
extrapolating these discrete values at finite $Q^2$ 
to $0$. 

However, there are potential problems in the 
$Q^2$ extrapolation, mainly due to the finite volume 
effects. On a finite lattice, the momentum transfer 
${\vec q}=(q_1,q_2,q_3)$ is quantized: $q_i = 2\pi \cdot n/N_i$ in unit of inverse 
lattice spacing, where $N_i$ is the total number of sites in
momentum direction $i$ and $n = 0, 1, ..., N_i$ is an integer. 
The smallest possible non-zero
$Q^2$ is equal to $-q_0^2+(2\pi/{\rm Max}(N_i))^2$, 
where $q_0 = \sqrt{M^2+ (2\pi/{\rm Max}(N_i))^2}-M$,
and $M$ is the nucleon mass. For a lattice with a spatial dimension
$16^3$, $Q^2_{\rm min} = 2M(\sqrt{(\pi/8)^2+M^2}-M)$. 
For $\beta=6$, this corresponds to $ \sim 0.5~
{\rm GeV}^2$. At this scale, the nucleon form factors 
are very sensitive functions of $Q^2$. Indeed $G_E(Q^2)$ 
increases from $Q^2=0.5~ {\rm GeV}^2$ to $Q^2=0$ by 
almost a factor of 3.
Therefore, an extrapolation across this range of
$Q^2$ must be strongly model dependent if without 
any prior knowledge of the $Q^2$ dependence of the
nucleon form factors 
\cite{wilcox92,kfl94,flynn96,flynn97}.
Fortunately, the experimental data shows that the 
$Q^2$-dependence of the nucleon form factors
can be fitted to a dipole form. Moreover, in a 
limited $Q^2$ range, the dipole form is not too 
different from a monopole. Therefore, we find in 
the literature both forms are commonly used to
fit the lattice data \cite{wilcox92,flynn97,capitani}. 
An example of the extrapolation using different 
fitting formula can be found in Ref. \cite{wilcox92}. 



In this paper, we report a more direct calculation of 
the nucleon magnetic moment using the elementary 
definition in electromagnetism \cite{jackson}. 
Although this approach was first mentioned in Ref. \cite{wilcox92}, 
to our knowledge, however, no systematic study along this line
has ever been reported in the literature. As is quite
obvious, the great advantage of this method is 
that it is free from the extrapolation of the finite 
momentum transfer $Q^2$. We will comment on a different
finite volume effect later in this paper.

Using a similar approach, we can calculate 
the fraction of the nucleon spin carried by the
quark angular momentum. In recent years, the study
of the spin structure of the nucleon has stimulated
much interest in both the experimental and theoretical nuclear
and particle community \cite{review,negele}. It was found in 
Ref. \cite{ji} that the quark angular momentum in the nucleon can
be obtained through deeply-virtual Compton scattering
and other hard exclusive processes. While this 
observation has stimulated 
much perturbative QCD and phenomenological study \cite{dvcs},
the quark angular momentum can 
also be calculated in lattice QCD. 
A first attempt in which the momentum
extrapolation was used in the magnetic moment
calculation was reported 
in Ref. \cite{mathur}. As shown in Ref. \cite{ji},
it can be calculated using the direct theoretical 
definition as well. In this paper, 
we also report such a calculation.

Our paper is organized as follows. 
In Sec. II, we give a brief theoretical discussion of 
the nucleon magnetic moment and quark angular momentum
in the continuum. We state our conventions and 
describe the implementation of the continuum formulas 
in the lattice simulations. 
In Sec. III, we outline the setup of our lattice
calculations and present our main results. We will 
compare them with previous calculations. 
We also consider the contributions of 
the disconnected diagrams to both 
the magnetic moment and quark total angular momentum. 
In the final section, we present the conclusion 
and discuss possible ways to improve upon the present 
calculation.

\section{Theoretical Details}

In this section, we present a number of theoretical
formulas which are useful in our numerical calculations. 
In the process, we will make clear our notations and 
conventions.  

\subsection{Magnetic Moment and Quark Angular Momentum}

According to the standard electromagnetic theory the 
magnetic moment operator of a system can be
defined as \cite{jackson}:
\begin{equation}
  {\vec \mu} = {1\over 2}\int [{\vec r} \times {\vec j}_{\rm em}] d^3r \ , 
\end{equation}
where $\vec{j}_{\rm em}$ is the 
electromagnetic current density. Its matrix element 
in a quantum state $|jm_z=j\rangle$
($\langle jm|jm\rangle=1$) quantized in the 
$z$-direction defines the magnetic moment,
\begin{equation}
  \mu = \langle jm_z=j|\hat \mu_z|jm_z=j\rangle  \ . 
\end{equation}
We will use this definition to calculate
the magnetic moment of the nucleon in its 
rest frame.
 
The magnetic moment can also be obtained from the 
form factors of the electromagnetic current. For the
nucleon, we have \cite{bjorkin}
\begin{equation}
\langle P' S' |j_{\rm em}^\mu(0)| PS \rangle =
 \overline{u}(P'S') \left[\gamma^\mu
F_1(Q^2) + i\frac{\sigma^{\mu\nu}q^\nu}{2m_N} F_2(Q^2)\right] u(PS) \ ,
\end{equation}
where $\sigma^{\mu\nu} = \frac{i}{2}[\gamma^\mu,\gamma^\nu]$, $|PS \rangle$ represents the ground state
of the nucleon with four-momentum $P$ and
polarization $S$, $q = P' - P$ is the momentum transfer, 
$Q^2 =- q_0^2 +{\vec q}^2$, and $u(PS)$ is an on-shell 
Dirac spinor. $F_1(Q^2)$ and 
$F_2(Q^2)$ are the well-known Dirac and Pauli form factors. 
The nucleon magnetic moment can be obtained 
from Sach's magnetic form factor 
\begin{equation}
G_M(Q^2) = F_1(Q^2) + F_2(Q^2)
\end{equation}
at $Q^2=0$: $\mu_N=G_M(0)$. This approach has been used 
in most of the lattice QCD calculations.

To understand the spin structure of the nucleon, 
one needs the QCD angular momentum operator in a 
gauge-invariant form 
\cite{ji,negele}
\begin{equation}
{\vec J}_{\rm QCD} = {\vec J}_{\rm g} + {\vec J}_{\rm q} \ ,
\end{equation}
with
\begin{eqnarray}
{\vec J}_{\rm q} &=& \int d^3r [{\vec r} \times {\vec T}] = 
\int d^3r \left[ \psi^\dagger
\frac{{\vec \Sigma}}{2} \psi + \psi^\dagger [{\vec r} \times (-i\overrightarrow{D})] 
\psi \right] \nonumber \ ,  \\
{\vec J}_{\rm g} &=& \int d^3r \left[ {\vec r} \times [{\vec E} \times {\vec B}] \right] \ ,
\end{eqnarray}
where ${\vec \Sigma}$ is the Dirac spin matrix 
and $\overrightarrow{D} = \overrightarrow{\nabla} 
- ig{\vec A}$ is the covariant derivative, ${\vec T}=(T^{j0})$ 
is a component of the quark energy-momentum tensor to be 
defined later. In a nucleon state with helicity 1/2
and momentum along the $z$-direction, the total helicity 
can be calculated as the expectation value of $J_z$, 
\begin{equation}
\frac{1}{2} = \frac{1}{2} \Sigma +  L_{\rm q} +  J_{\rm g} 
= J_{\rm q} + J_{\rm g} \ ,
\end{equation}
where the various contributions
to the spin of the nucleon are defined as
the expectation values of the corresponding 
operators. For instance,
the matrix element of ${\vec J}_{\rm q}$
defines the quark total angular momentum contribution 
to the nucleon spin. 

As in the case of the nucleon magnetic moment, 
the quark contribution to the nucleon spin $J_q$
can be obtained from the form factors of the 
quark energy-momentum tensor $T^{\mu\nu} = \frac{i}{2}{\overline \psi}
\stackrel{\leftrightarrow}{D^{(\mu}} \gamma^{\nu)} \psi$ (the
parentheses around the two indices means symmetrization
and subtraction of the trace and $\stackrel{\leftrightarrow}D
=\stackrel{\rightarrow}D-\stackrel{\leftarrow}D$) 
\begin{eqnarray}
&& \langle P'S' |T^{\mu\nu}(0)| PS \rangle  = \nonumber \\
&&
\overline{u}(P'S') \left [ \overline{P}^{(\mu} \gamma^{\nu)}
T_1(Q^2) + \frac{\overline{P}^{(\mu}i\sigma^{\nu)\beta}
        q^\beta}{2M} T_2(Q^2) + \frac{q^{(\mu}q^{\nu)}}{M} T_3(Q^2)  \right] u(PS) \ ,
\end{eqnarray}
where $\overline{P}^\mu = (P^\mu + P'^\mu)/2$.
The quark contribution to the spin of the nucleon
is then simple $J_{\rm q} = \frac{1}{2} [T_1(0) + T_2(0)]$. 
In Ref. \cite{mathur}, this was used to calculate $J_q$.

\subsection{Lattice Formalism}

Light (up and down) quarks are put on the lattice using the standard 
Wilson formulation. In the discrete Euclidean spacetime, 
the fermionic part of the QCD action is given by \cite{wilson}:
\begin{equation}
S_W  = \sum\limits_{i,j} \overline{\psi}_i M_{ij} \psi_j  \ , 
\end{equation}
where $i, j$ is the sum over the full set of spacetime, Dirac, color indices
with $i=\{x,\alpha,a\}$ and $j=\{y,\beta,b\}$. $M_{ij}$ is defined as
\begin{equation} 
M_{ij} = \delta_{i,j} - \kappa \sum\limits_\mu \left\{
(1-\gamma_\mu)_{\alpha\beta} U_\mu(x)^{ab} \delta_{x,y-\hat \mu} + 
(1+\gamma_\mu)_{\alpha\beta} U^\dagger_\mu(x-\hat \mu)^{ab}
\delta_{x,y+\hat \mu} \right\} \ , 
\end{equation}
where $\kappa = 1/(2(m_0+4))$ with $m_0$ the bare quark mass, 
$\hat \mu$ is a unit vector along the $\mu=1,2,3,4$ direction, 
and the lattice spacing $a$ is set to 1. 
$U_\mu(x) = e^{iA_\mu(x)}$ represents a link connecting 
the nearest lattice sites going from $x$ to $x+\hat \mu$. 
Our notation for $\gamma$-matrices follows Refs. \cite{rothe,andrew}:
\begin{equation}
\gamma_4^E = \gamma^0 = \pmatrix{ I & 0\cr 0 & -I}, \hspace{3mm}
\gamma^E_i = -i \gamma^i 
= \pmatrix{0 & -i \sigma_i\cr i \sigma_i & 0} \ , 
\end{equation}
where $E$ means Euclidean space and will be omitted henceforth. 
The standard continuum Dirac field $\psi$ is related to 
the above by an extra factor of $\sqrt{m_0+4}$. 

\begin{figure}[t]
\begin{center}
\epsfig{figure=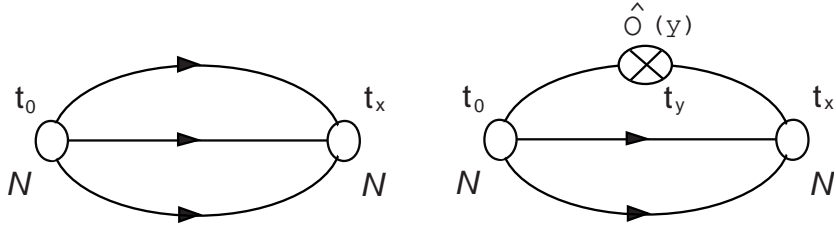,height=4cm,clip=,angle=0}
\caption{Two-point and three-point diagrams.}
\label{fig:1}
\end{center}
\end{figure}

We are interested in calculating two- and three-point 
correlation functions involving the nucleon 
interpolating fields on the lattice. The graphical 
representation for them is 
shown in Fig.\ref{fig:1}, where $t_0$ is the ``time" of the 
nucleon creation (source position), 
$t_x$ is the ``time" of the nucleon annihilation 
(sink position) and $t_y$ is 
the ``time" of the operator insertion (we sum only
over the spatial coordinates ${\vec x}$ and ${\vec y}$).

The nucleon two-point correlation function is 
defined as 
\begin{equation}
G_{2\alpha\alpha'}(t_x,{\vec P'})  = \sum\limits_{\vec x} e^{-i{\vec P'} {\vec x}} \langle 0
|\chi_\alpha(x) \overline{\chi}_{\alpha'}(0) | 0 \rangle \ , 
\end{equation}
where $x=(t_x,{\vec x})$ are discretized spacetime
coordinates, $\alpha, \alpha'$ are Dirac indices, and $|0 \rangle$ 
represents QCD vacuum. Operator $\chi$ is an interpolating 
field with the nucleon quantum number. 
We choose \cite{gupta,ashman,leinweber93}
\begin{equation}
\chi_\alpha = \epsilon^{ijk} (u^iC\gamma_5d^j)u^k_\alpha \ . 
\end{equation}
Indices $\{i,j,k\} = 1,2,3$ represent colors, $C = \gamma_4 \gamma_2$ is a
charge-conjugation matrix. The time dependence of the 
two-point function can be obtained by inserting a 
complete set of intermediate states $|n\rangle$
with the nucleon quantum number: 
\begin{equation}
G_{2\alpha\alpha'}(t_x,{\vec P'}) = \sum\limits_n  
\langle 0|\chi_\alpha(0)|n \rangle 
\langle n|\overline \chi_{\alpha'}(0)
|0 \rangle V_3 e^{-E_n t_x} \ ,
\end{equation}
where we have used the Euclidean translation 
$\chi(\vec x,t_x) = e^{(Ht_x-i\vec x \vec p)} 
\chi(0) e^{(-Ht_x+i\vec x \vec p)}$ and $V_3$ is the 
three volume. In the limit $t_x>\!\!>1$, only the lightest 
state (namely, the nucleon) with momentum $\vec{P'}$ 
contributes. Therefore the nucleon mass can 
be extracted from the (large) time dependence of 
the two-point function, 
\begin{equation} \label{masseq}
M = {\rm ln} \frac{G_{2\alpha\alpha}(t_x,{\vec P'}=0)}
{G_{2\alpha\alpha}(t_x+1,{\vec P'}=0)} \ .
\end{equation}
$M$ is in the unit of the inverse lattice spacing.
In the next section, we will extract the nucleon mass
this way as it is needed for the magnetic moment
calculation.

The matrix element of an observable $\widehat O$ 
in the nucleon state can be extracted from the 
following three-point function: 
\begin{equation}
G_{3\alpha\alpha'}(t_x, t_y, {\vec P}, {\vec P'})  
= \sum\limits_{\vec x, \vec y} e^{-i {\vec P'} \vec x}
e^{i {\vec q} {\vec y}} \langle 0 |\chi_\alpha(x) 
\widehat{O}(y)  \overline{\chi}_{\alpha'}(0) |0 \rangle  \ . 
\end{equation}
At large time separation, $t_0 <\!\!< t_y <\!\!< t_x$, 
it can be written as 
\begin{eqnarray}
&& G_{3\alpha\alpha'}(t_x, t_y, {\vec P}, {\vec P'}) = \nonumber \\
&& V_3^2\sum\limits_{S,S'} \langle 0|\chi_\alpha(0)|P'S'\rangle
\langle P'S'| \widehat{O}(0) |PS \rangle \langle PS|\overline{\chi}_{\alpha'}(0)|0 \rangle
\cdot e^{-E_{P'}t_x} \cdot e^{-(E_P-E_{P'})t_y} \ ,
\end{eqnarray}
where we have kept only the intermediate nucleon
state. The labels $S$ and $S'$ denote the nucleon polarization.
The nucleon state here is assumed to be normalized to 1. 

In this paper, we are interested in 
the electromagnetic current operator 
$j^{\rm em}_\mu(y)$ and quark energy-momentum tensor 
operator $T_{\mu\nu}(y)$. The continuum expression 
$j^{\rm em}_\mu(y) = \sum_i e_i \bar\psi_i\gamma_\mu \psi_i$ 
(where $\psi$ is the quark field of flavor $i$ and $e_i$ is the
electric charge) is a conserved current \cite{karsten,martinelli}. 
On a lattice, the above expression can be implemented straightforwardly, but 
it is not conserved because of the finite lattice 
spacing. Instead, $j^{\rm em}_\mu$ is multiplicatively 
renormalized if the power-suppressed contributions are
neglected. The conserved electromagnetic current 
in the Wilson fermion formalism has the following
more complicated form
\cite{martinelli_pion}:
\begin{equation}
j^{\rm cons}_\mu(y) = \frac{1}{2} \left[
\overline{\psi}(y) (\gamma_\mu - 1) U_\mu(y) 
\psi(y+\hat \mu) + \overline{\psi}(y+\mu)(\gamma_\mu + 1) 
U^\dagger_\mu(y)
\psi(y)\right] \ . 
\end{equation}
Differences between local and conserved currents 
on a lattice are discussed in Refs. \cite{martinelli,martinelli_pion}.

The quark energy-momentum tensor 
$T_{\mu\nu}(y)$ is not conserved. Therefore, 
there is no preferred way to put the operator
on a lattice. We choose the local form, 
$T^{\mu\nu}(y) = \frac{i}{2} \overline{\psi}(y) \{ \gamma^\mu
[\overrightarrow{D} - \overleftarrow{D}]^\nu \} \psi(y)$ as in the continuum 
with the derivative $\overrightarrow{D}_\mu$ defined on a 
lattice as:
\begin{equation}
\overrightarrow{D}_\mu \psi(x) = \frac{1}{2} [U_\mu(x) \psi(x+\hat\mu) -
U^\dagger_\mu(x-\hat \mu) \psi(x-\hat \mu)] \ . 
\end{equation}
The above definition is related to 
the continuum one, say, the $\overline{MS}$ scheme,
through a finite renormalization. Ignoring the  
mixing contribution from the gluon energy-momentum 
tensor, the tadpole improved renormalization constant
for operator $T_{\mu\nu}$ has been calculated 
perturbatively and is near unity, $Z = 1.045$
\cite{capitani97}. A nonperturbative renormalization
technique may be needed to find the renormalization
factor reliably. 

Finally, we consider the transformation of 
operators from the Minkowski to Euclidean space.
When the vector current $\overline{\psi}\gamma_\mu\psi$
is defined with the Euclidean $\gamma$ matrices, 
the corresponding Lorentz vectors in Eq. (3), 
for example, must be defined with $V_4 = V^0$
and $V_i = -i V^i_M$. Alternatively, one can
define the current as $\overline{\psi}i\gamma_\mu\psi$, 
then the Euclidean four-vectors are related to 
the Minkowski ones by $V_4=iV^0$ and $V_i = V^i_M$. 

\subsection{Formulas for Physical Observables on Lattice}

As we have discussed earlier, the nucleon magnetic moment and quark total angular momentum can
be calculated by an extrapolation of the relevant form factors 
to $Q^2=0$ \cite{draper}-\cite{dong}. On a lattice, 
these form factors can be extracted from the ratio of
three- to two-point correlation functions. 
For the magnetic form factor $G_M(Q^2) = F_1(Q^2) + F_2(Q^2)$ and the total angular momentum
$T(Q^2) = \frac{1}{2} [T_1(Q^2) + T_2(Q^2)]$, one can obtain at the large time separation, 
\begin{equation}
\frac{\sum\limits_{\vec x,\vec y} e^{i \vec q \vec y - i \vec {P'} \vec x}
\langle 0 | \chi_\alpha(x) J_i(y) \overline \chi_{\alpha'}(0)
| 0 \rangle}
{\sum\limits_{\vec x} e^{- i \vec {P'} \vec x} \langle 0
| \chi_\gamma(x) \overline \chi_{\gamma'}(0) | 0 \rangle}
=
\frac{f}{(\hat {P'} + M)_{\gamma\gamma'}} \epsilon^{ijk} \sigma^j_{\alpha\alpha'} q^k G_M(Q^2) \ , 
\end{equation}

\begin{equation}
\frac{\sum\limits_{\vec x,\vec y} e^{i \vec q \vec y - i \vec {P'} \vec x}
\langle 0 | \chi_\alpha(x) T_{4i}(y) \overline \chi_{\alpha'}(0)
| 0 \rangle}
{\sum\limits_{\vec x} e^{- i \vec {P'} \vec x} \langle 0
| \chi_\gamma(x) \overline \chi_{\gamma'}(0) | 0 \rangle}
= \frac{\overline{P}^0}{i} \frac{f}{(\hat {P'} + M)_{\gamma\gamma'}} \epsilon^{ijk}
\sigma^j_{\alpha\alpha'} q^k T(Q^2) \ ,
\end{equation}
where $\sigma$ is the Pauli matrix, and the multiplier $f$ is equal to 
$$
f = e^{-E_{P}t_y} \cdot e^{E_{P'}t_y} = \left [ \frac{E_{P'} + M}
{2E_{P'}} \right ]^{\frac{t_y}{t_x}} \cdot \left [ \frac{2E_{P}}{E_{P} + M}
\right ]^{\frac{t_y}{t_x}} \cdot \left [ \frac{
\sum\limits_{\vec x} e^{- i \vec {P} \vec x} \langle 0
| \chi_\alpha(x) \overline \chi_{\alpha'}(0) | 0 \rangle}
{\sum\limits_{\vec x} e^{- i \vec {P'} \vec x} \langle 0
| \chi_\alpha(x) \overline \chi_{\alpha'}(0) | 0 \rangle} \right ]^{\frac{t_y}{t_x}} \ ;
$$
$f = 1$ when $\vec q = 0$. These expressions are used 
in this study to 
calculate form factors to check against the existing results. 

The nucleon magnetic moment and quark total angular momentum can be calculated by taking
the derivative over the momentum transfer in the limit $\vec q \rightarrow 0$:

\begin{eqnarray}
&& \frac {\sum\limits_{\vec x,\vec y} \langle 0 | \chi_\alpha(x) y_j J_i(y)
\overline \chi_{\alpha'}(0) | 0 \rangle}
{\sum\limits_{\vec x} \langle 0 | \chi_\gamma(x) \overline \chi_{\gamma'}(0) | 0 \rangle} 
 = 
\frac{\partial}{\partial(iq_j)} \frac{\sum\limits_{\vec x,\vec y}
e^{i \vec q \vec y - i \vec {P'}
\vec x} \langle 0 | \chi_\alpha(x) J_i(y) \overline \chi_{\alpha'}(0) | 0 \rangle}
{\sum\limits_{\vec x} e^{-i{\vec P'} {\vec x}} \langle 0
|\chi_\gamma(x) \overline \chi_{\gamma'}(0) | 0 \rangle} \nonumber  \\
&& = \frac{\partial}{\partial(iq_j)} \frac{f}{(\hat {P'} + M)_{\gamma\gamma'}}
\epsilon^{ikj} \sigma^k_{\alpha\alpha'} q^j G_M(Q^2) \stackrel{\vec q \rightarrow 0}{=} 
\frac{i \epsilon^{ijk} \sigma^k_{\alpha\alpha'}}{2M} \mu_N  \ ,
\end{eqnarray}

\begin{eqnarray}
&& \frac{ \sum\limits_{\vec x,\vec y} \langle 0 | \chi_\alpha(x) y_j T_{4i}(y)
\overline \chi_{\alpha'}(0) | 0 \rangle}
{\sum\limits_{\vec x} \langle 0 |\chi_\gamma(x) \overline \chi_{\gamma'}(0) | 0 \rangle} = 
\frac{\partial}{\partial(iq_j)} \frac{\sum\limits_{\vec x,\vec y} 
e^{i \vec q \vec y - i \vec {P'} \vec x}
\langle 0 | \chi_\alpha(x) T_{4i}(y) \overline \chi_{\alpha'}(0) | 0 \rangle}
{\sum\limits_{\vec x} e^{-i{\vec P'} {\vec x}} \langle 0
|\chi_\gamma(x) \overline \chi_{\gamma'}(0) | 0 \rangle} \nonumber  \\
&& = \frac{\partial}{\partial(iq_j)} \frac{i f \overline P^0}{(\hat {P'} + M)_{\gamma\gamma'}}
\epsilon^{ijk} \sigma^k_{\alpha\alpha'} T(Q^2) \stackrel{\vec q \rightarrow 0}{=}
\frac{\epsilon^{ijk} \sigma^k_{\alpha\alpha'}}{2} J_q \ .
\end{eqnarray}


The summations over $\vec{x}$ and $\vec{y}$
ensure the nucleon has vanishing three-momentum (in the
rest frame) and the forward matrix elements are 
calculated. Therefore the Dirac index $\gamma'$
must be the same as $\gamma$. The nucleon can be polarized
in the different spatial direction by selecting different
$\alpha$ and $\alpha'$. The new results obtained 
from the above formulas will be presented in the next section.

\section{Numerical Results}

In this section, we present our 
numerical calculations on the $16^3 \times 32$ 
lattice at $\beta = 6.0$ with the Wilson
formulation of fermions. 
The results include the nucleon mass, form factors,
magnetic moment, and the quark total angular momentum.

The coupling ($\beta$) corresponds to lattice spacing 
$a \sim 0.1~{\rm fm}$ or $a^{-1} \simeq (1.7-1.9) ~{\rm GeV}$.
The antiperiodic boundary condition for the Dirac operator
is used in the time direction. We use  
100 quenched configurations at every quark mass
to attain reasonable statistics. However, as the quark 
masses approach the chiral limit ($\kappa$ reaches
$\kappa_{\rm cr} \simeq 0.1568(1)$ \cite{mathur}), 
the statistics require many more lattice configurations, 
and hence the error bars increase markedly for a fixed
number of configurations. 

We use three different values of the mass parameter
$\kappa = 0.152, ~0.154, ~0.155$, corresponding to 
quark masses $210, ~124, ~80~ {\rm MeV}$, respectively. 
Point sources are used to create the nucleon: 
interpolating field $\overline\chi_\alpha$ 
is placed on a well-defined initial 
lattice point, called 0, and twelve different sources
with different color and Dirac indices 
are used to start the conjugation gradient.

Errors are determined by the standard 
jackknife procedure \cite{yang}-\cite{shao}. 
We average the three-point and two-point functions separately 
using this method for 100 configurations
and put already averaged values in the final formulas for the magnetic moment
and quark total angular momentum.

\subsection{Nucleon Mass}

\begin{figure}[t]
\begin{center}
\epsfig{figure=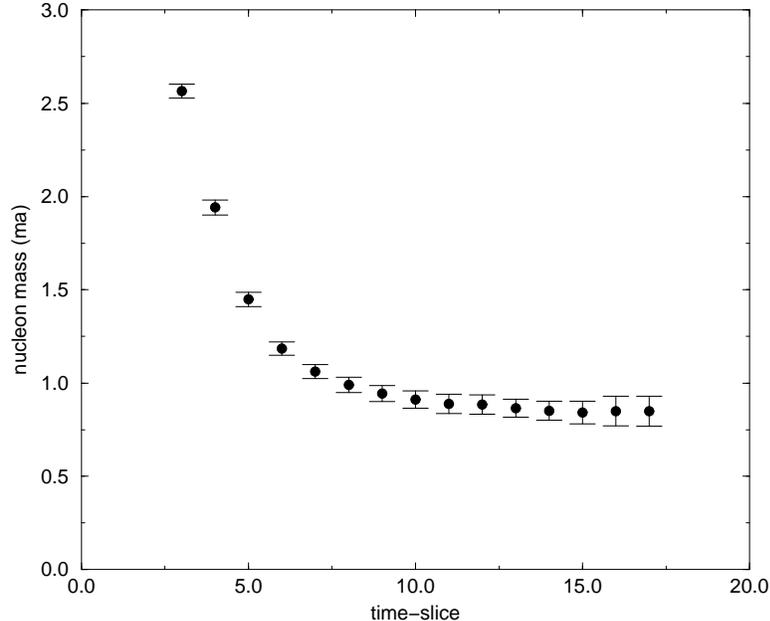,width=10cm,clip=,angle=-90}
\caption{Nucleon mass as a function of time-slice, calculated
on $16^3\times 32$ lattice with 100 quenched gauge configurations at $\beta=6.0$ and
$\kappa = 0.152$.}
\label{fig:2}
\end{center}
\end{figure}

As is clear from the formulas in the last
section, we need the lattice nucleon mass to calculate
the nucleon electromagnetic form factors and magnetic 
moment. We use Eq.$\!$\ref{masseq} to extract the nucleon mass
on the lattice. The result is shown in Fig.$\!$\ref{fig:2} as a 
function of the time-slice for $\kappa = 0.152$. 
A stable plateau is seen to have reached 
between the time slices 10 and 17. 

Table \ref{mass} shows the comparison between the 
nucleon masses obtained in our calculations 
and those in the literature (Ref. \cite{wilcox92,mathur}).
The results, shown in the units of the inverse
lattice spacing $a^{-1}$, are in good agreement. 
Note that the lattice nucleon masses are
consistently bigger than the 
physical mass of $\sim 0.534$ in 
the lattice unit because of the large lattice 
quark masses.

\begin{table} 
\caption{Nucleon mass in lattice units ($ma$), calculated
on $16^3\times 32$ lattice with 100 quenched gauge configurations at $\beta=6.0$}
\begin{tabular}{|c|c|c|c|}  
\hspace{3mm} $\kappa$  \hspace{3mm} &  \hspace{5mm}  nucleon mass
\hspace{5mm} & \hspace{3mm} nucleon mass from Ref.\cite{wilcox92} \hspace{3mm} &
 \hspace{3mm} nucleon mass from Ref.\cite{mathur} \hspace{3mm} \\ 
\hline \hline
              &                 &           & \\  
 0.152     &   0.86(2) &   0.87(3)  &  0.882(12)   \\  
\hline
              &                       &           & \\  
 0.154     &   0.7(3) &  0.73(5) &  0.738(16)   \\  
\hline
              &                          &           &  \\  
0.155  &  0.63(5)  &          &   0.67(15) \\  
\end{tabular} 
\label{mass}
\end{table}

\subsection{Nucleon Magnetic Moment and Quark Total Angular Momentum}

We first present the results for the
nucleon form-factors at several momentum transfers 
$Q^2$ and compare them 
with the previous calculations \cite{wilcox92}.  
The expected consistency is a useful check 
of the codes for three-point functions. 
Results are compared in Tables \ref{ff1}
(for ${\vec q}^2=0.154$) and \ref{ff2} (for ${\vec q}^2=0.308$) 
and are seen in agreement within error bars.
The agreement improves for larger $\kappa$. One 
noticeable trend is that when the quark mass
becomes smaller, the magnetic form factor decreases. 
This could be an indication 
that the finite lattice size effect is important.
We will comment on this further below. 

\begin{table} 
\caption{Magnetic form-factors $G_M(q^2)$ for ${\vec q}^2 a^2 = 0.154$, calculated
on $16^3\times 32$ lattice with 100 quenched gauge configurations at $\beta=6.0$}
\begin{tabular}{|c|c|c|c|c|}  
\hspace{3mm} $\kappa$  \hspace{3mm} &  \hspace{5mm}  proton
\hspace{5mm} & \hspace{5mm} neutron \hspace{5mm} &
\hspace{3mm} proton from Ref.\cite{wilcox92} \hspace{3mm} &
 \hspace{3mm} neutron from Ref.\cite{wilcox92} \hspace{3mm} \\ 
\hline \hline
              &               &            &           & \\  
 0.152     &   1.36(9) &  -0.88(7) &  1.22(7) & -0.781(59)   \\  
\hline
              &               &              &           &  \\  
0.154  &  1.13(12)  &  -0.738(75)  &  1.17(11) &   -0.748(69) \\  
\hline
              &               &              &           &  \\  
0.155  &  1.02(15)  &  -0.661(101)  &   &    \\  
\end{tabular} 
\label{ff1}
\end{table}

\begin{table} 
\caption{Same as Table \ref{ff1}, but for ${\vec q}^2 a^2 = 0.308$, calculated
on $16^3\times 32$ lattice with 100 quenched gauge configurations at $\beta=6.0$}
\begin{tabular}{|c|c|c|c|c|}  
\hspace{3mm} $\kappa$  \hspace{3mm} &  \hspace{5mm}  proton
\hspace{5mm} & \hspace{5mm} neutron \hspace{5mm} &
\hspace{3mm} proton from Ref.\cite{wilcox92} \hspace{3mm} &
 \hspace{3mm} neutron from Ref.\cite{wilcox92} \hspace{3mm} \\ 
\hline \hline
              &               &            &           & \\  
 0.152     &   1.09(8) &  -0.697(65) &  0.906(59) & -0.586(47)   \\  
\hline
              &               &              &           &  \\  
0.154  &  0.877(97)  &  -0.567(93)  &  0.895(90) &   -0.578(89) \\  
\end{tabular} 
\label{ff2}
\end{table}				

\begin{table} 
\caption{Nucleon magnetic moments, calculated
on $16^3\times 32$ lattice with 100 quenched gauge configurations at $\beta=6.0$}
\begin{tabular}{|c|c|c|}  
              &                  &     \\  
\hspace{5mm} $\kappa$ \hspace{5mm} & proton \hspace{20mm} &  neutron  \hspace{20mm} \\ 
\hline \hline
              &               &          \\  
\hspace{5mm} 0.152  \hspace{5mm}   & 2.24(19) \hspace{20mm} &  -1.46(13) \hspace{20mm}  \\  
\hline
              &               &           \\  
\hspace{5mm} 0.154 \hspace{5mm} &  2.09(30) \hspace{20mm} & -1.38(21) \hspace{20mm} \\  
\hline
              &               &       \\  
\hspace{5mm} 0.155 \hspace{5mm} & 1.88(37) \hspace{20mm}  &  -1.26(28) \hspace{20mm} \\  
\end{tabular} 
\label{mm}
\end{table}

\begin{table} 
\caption{Quark total angular momentum, calculated
on $16^3\times 32$ lattice with 100 quenched gauge configurations at $\beta=6.0$}
\begin{tabular}{|c|c|c|}  
              &        &        \\  
\hspace{10mm} $\kappa$ \hspace{10mm} & \hspace{10mm} connected $J_{\rm u+d}$ \hspace{10mm}
& \hspace{10mm} disconnected $J_{\rm u+d}$ \hspace{10mm}  \\  
\hline \hline
              &      &          \\  
\hspace{10mm} 0.152 \hspace{10mm} & \hspace{10mm} 0.47(7) \hspace{10mm} & \hspace{10mm}
-0.12(6) \hspace{10mm} \\  
\hline
              &       &        \\  
\hspace{10mm} 0.154 \hspace{10mm} & \hspace{10mm} 0.45(9) \hspace{10mm} & \\  
\hline
              &         &      \\  
\hspace{10mm} 0.155 \hspace{10mm} & \hspace{10mm} 0.44(10) \hspace{10mm}  & \\  
\end{tabular}
\label{tam}
\end{table}

Results for the nucleon magnetic moment are summarized in 
Table \ref{mm}. One obvious conclusion from the table is that the lattice
magnetic moments are about 30\% smaller than 
the experimental data ($\mu_p = 2.79$ and $\mu_n = -1.91$). 
There are a number of factors which can account for
this discrepancy. One is the use of the quenched 
approximation. We will report a full dynamical
calculation in a future publication. 
Another is the lattice quark masses.  
The smallest quark mass in our calculations 
is $80~{\rm MeV}$  for $\kappa  = 0.155$, which 
corresponds to the pion mass $\sim 0.5~{\rm GeV}$. 
In a phenomenological study \cite{leinweber99,leinweber01}, 
Leinweber et. al. demonstrated that as the pion mass 
approaches zero, the magnetic moment shows a steep rise 
as a result of chiral singular contributions.
In our calculations, we observe an opposite 
trend: the magnetic moment decreases along
with the lattice quark masses. This is seen in 
the calculation of the form factors at finite momentum
transfers and a number of other observables \cite{bnl}.
We suspect strongly that this is a finite volume effect
which will go away when we increase the physical
dimension of the lattice by 50\% or so. 

The results for the quark total angular momentum contribution
to the nucleon spin are shown in Table \ref{tam}.
In this case, our results are quite consistent with
those in Ref. \cite{mathur}, although we have avoided
the use of the extrapolation in $Q^2$. For instance,
our result $J_{\rm u+d} = 0.44(10)$ at
$\kappa = 0.155$ from the connected diagram
can be compared with $J_{\rm u+d} = 0.44(7)$ at
$\kappa_{\rm cr} = 0.1568$ in \cite{mathur}. 
The leading chiral contribution to this quantity 
was studied in Ref. \cite{chenji} and is 
not strong. As the quark mass approaches 
the chiral limit, the lattice result shows a 
slight decrease, although the large 
error bars prevent a clear-cut comparison. 

Although the quark angular momentum from the
connected diagram accounts for nearly all of 
the spin of the nucleon, there are other contributions
which cannot be neglected. For instance, the
sea contribution through the disconnected diagrams
is expected to quench the connected result, resulting
a total quark angular momentum accounting for
about half of the nucleon spin \cite{ji2}. 

\subsection{Disconnected Diagrams}

\begin{figure}[t]
\begin{center}
\epsfig{figure=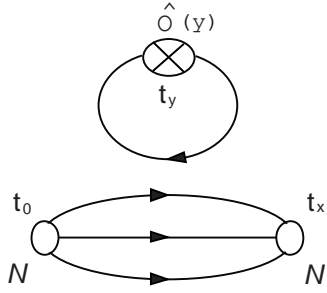,height=4cm,clip=,angle=0}
\caption{Disconnected diagram.}
\label{fig:3}
\end{center}
\end{figure}

The contribution of the disconnected diagrams represents
one of the possible corrections to the nucleon 
magnetic moment and quark total angular momentum presented
in the previous subsection.
The disconnected insertion is shown in Fig.$\!$\ref{fig:3}. 
Disconnected diagrams are difficult to
calculate in general because they involve the 
closed-loop contributions resulting 
from the self contractions of the quark fields
in an observable. An efficient method 
to obtain the matrix elements is the 
noise theory discussed in Refs.\cite{thron}-\cite{mathur01}, 
explained as follows.

Consider a set of random sources $\eta^l(x)$ 
satisfying
\begin{equation}
\frac{1}{N_{\rm total}} \sum\limits_{l=1}^{N_{\rm total}} 
\eta^{\dagger l}(x) \eta^l(y) =
\delta_{xy} \ . 
\end{equation}
The total number of sources $N_{\rm total}$ 
is a key parameter controlling the desired statistics. 
In this work, the random sources $\eta^l(x)$ are 
created by putting $1,~ -1,~ i$
or $-i$ randomly on every lattice point for every 
Dirac and color index. Three-point functions for 
the disconnected diagrams can now be expressed as:
\begin{equation}
G_{3\alpha\alpha'}^{\rm disc}(t_x, t_y, {\vec P'}) = G_{2\alpha\alpha'}(t_x, {\vec P'}) \cdot
\frac{1}{N_{\rm total}}
\sum\limits_{l=1}^{N_{\rm total}} \sum\limits_{{\vec z,\vec y}}
\eta^{\dagger l}(y) [{\vec y} \times \widehat{O}(y)] M^{-1}(y,z) \eta^l(z) \ , 
\end{equation}
where $\widehat{O} = \vec {j}^{\rm cons}$ or $T^{j0}$, 
$G_{2\alpha\alpha'}(t_x, {\vec P'})$ is Green's two-point function,
and $M^{-1}(y,x)$ is a Wilson-fermion propagator. 

Unfortunately, this method requires a large $N_{\rm total}$
to obtain reasonable statistics, which is very 
expensive in practical simulations. It was discussed 
in Refs.\cite{wilcox01,mathur01} that the so-called expansion 
method can expedite the calculation significantly. 
The idea is to approximate the full fermion propagator, 
\begin{equation}
M^{-1}(x,y) = \frac{1}{\delta_{xy} - \kappa P_{xy}} \ ,  
\end{equation}
where $P_{xy} = \sum\limits_\mu [(1+\gamma_\mu)U_\mu(x)\delta_{x,y-\hat \mu} +
(1-\gamma_\mu)U^\dagger_\mu(x-\hat \mu)\delta_{x,y+\hat \mu}]$ by
one that includes only the first few terms in the small
$\kappa$ expansion, 
\begin{equation}
M^{-1}_{\rm exp}(x,y) = \delta_{xy} + \kappa P_{xy} + \kappa^2 P_{xy}^2 + ... \ . 
\end{equation}
One then writes $M^{-1}(x,y) = M^{-1}_{\rm exp}(x,y) + \{
M^{-1}(x,y) - M^{-1}_{\rm exp}(x,y) \}$. The first term, 
$M^{-1}_{\rm exp}(x,y)$, can be calculated directly
without conjugation gradient, and 
the remainder, $M^{-1}(x,y) - M^{-1}_{\rm exp}(x,y)$, 
is computed by the noise method discussed above.
Because the remaining contribution is small, the number of
random sources needed for the required statistics
is considerably reduced.

In this study, we keep ten terms in $M^{-1}_{\rm exp}$. 
The remainder is calculated with 30 random sources. 
The results are obtained only for the smallest 
$\kappa = 0.152$ because even in this case the 
disconnected contribution to the magnetic moment has 
almost $100\%$ error. For the quark total 
angular momentum, the result is more stable 
and the error is about $50\%$. Our final result
for the disconnected diagram is $J_{u+d} = -0.12(6)$. 
This agrees with \cite{mathur} $J_{u+d} = 
-0.094(24)$. The negative contribution 
cancels part of the result from the connected diagram. 
We have no simple explanation for the sign.

\section{Conclusion}

In this paper, we report
a lattice QCD calculation of the nucleon 
magnetic moment and the fraction of the nucleon
spin carried in the quark total angular momentum. 
We use the standard Wilson formulation for 
fermions and 100 quenched configurations in the
Monte Carlo evaluations of the 
Feynman path integrals.

Our results for the magnetic moment cannot be compared
directly with the experiment yet. We are uncertain,
for example, about the systematic error caused by 
the quenched approximation, large lattice quark masses, 
and the finite volume effect. The smallest quark mass
in our calculations is about $80 ~{\rm MeV}$
($\kappa = 0.155$) which is about ten times larger than 
the physical quark mass. The pion mass dependence
of the magnetic moment has been studied recently in 
Refs. \cite{leinweber99,leinweber01}. In these papers,
a strong dependence on the pion (quark) mass in the 
region ${m_\pi}^2 < 0.5~ {\rm GeV}$ is suggested. 
A phenomenological chiral extrapolation of the previous
lattice results to the physical
pion mass gives results consistent with experimental
values. 

One interesting finding of this study 
is that the magnetic form factors, magnetic moment,
and the quark total angular momentum 
decrease with the quark masses. The most
plausible explanation is that when the quark mass
becomes lighter, the nucleon size grows, and 
the present lattice volume cannot fully accommodate 
a physical nucleon.

We have also studied the contribution of the 
disconnected diagrams to various physical observables. 
The error bars grow very significant at small
quark masses. The sea contribution to the quark total angular 
momentum is found to be negative. The fraction
of the nucleon spin carried by the up and down quark
angular momentum is $J_q= 0.35\pm 10$ 
at $\kappa=0.152$. The result for a fully dynamical
simulation will be reported elsewhere.

\acknowledgments   
The numerical calculations reported here
were performed on the Calico Alpha Linux Cluster and the QCDSP
at the Jefferson Laboratory, Virginia.
C.J., X.J. and V.G. are supported in part by funds provided by the
U.S.  Department of Energy (D.O.E.) under cooperative agreement
DOE-FG02-93ER-40762. 
V.G. was supported in part by research fellowship from the Southeastern Universities
Research Association (SURA).










\end{document}